\documentclass[acmsmall]{acmart}

\AtBeginDocument{%
  \providecommand\BibTeX{{%
    \normalfont B\kern-0.5em{\scshape i\kern-0.25em b}\kern-0.8em\TeX}}}

\setcopyright{none}

\begin{document}

\title{Interacting with next-phrase suggestions: How suggestion systems aid and influence the cognitive processes of writing}

\author{Advait Bhat}
\email{advaitmb@gmail.com}
\orcid{0000-0002-5524-2387}
\affiliation{%
  \institution{Indian Institute of Technology Bombay}
  \city{Mumbai}
  \state{Maharashtra}
  \country{India}
}

\author{Saaket Agashe}
\email{saagashe@ucsc.edu}
\affiliation{%
  \institution{University of California Santa Cruz}
  \city{Santa Cruz}
  \state{California}
  \country{USA}
  }

\author{Niharika Mohile}
\author{Parth Oberoi}
\author{Ravi Jangir}
\affiliation{%
   \institution{Indian Institute of Technology Bombay}
  \city{Mumbai}
  \state{Maharashtra}
  \country{India}
}

\author{Anirudha Joshi}
\email{anirudha@iitb.ac.in}
\affiliation{%
   \institution{Indian Institute of Technology Bombay}
  \city{Mumbai}
  \state{Maharashtra}
  \country{India}
}

\renewcommand{\shortauthors}{Bhat, et al.}

\begin{abstract}
   Writing with next-phrase suggestions powered by large language models is becoming more pervasive by the day. However, research to understand writers' interaction and decision-making processes while engaging with such systems is still emerging. We conducted a qualitative study to shed light on writers' cognitive processes while writing with next-phrase suggestion systems. To do so, we recruited 14 amateur writers to write two reviews each, one without suggestions and one with suggestions. Additionally, we also positively and negatively biased the suggestion system to get a diverse range of instances where writers' opinions and the bias in the language model align or misalign to varying degrees. We found that writers interact with next-phrase suggestions in various complex ways: Writers abstracted and extracted multiple parts of the suggestions and incorporated them within their writing, even when they disagreed with the suggestion as a whole; along with evaluating the suggestions on various criteria. The suggestion system also had various effects on the writing process, such as altering the writer's usual writing plans, leading to higher levels of distraction etc. Based on our qualitative analysis using the cognitive process model of writing by \citet{doi:10.1177/0741088312451260} as a lens, we propose a theoretical model of 'writer-suggestion interaction' for writing with GPT-2 (and causal language models in general) for a movie review writing task, followed by directions for future research and design.
\end{abstract}


\maketitle

\section{Introduction}
With the advent of transformer-based probabilistic language models \cite{vaswani2017attention}, text generation has become sophisticated and viable enough to be used in writing interfaces to provide synchronous \emph{next-phrase suggestions}, as seen in popular products like Google Smart Compose \cite{chen2019gmail} and LightKey \cite{lightkey}. As the writer types, the editor predicts an in-line phrase that the writer can select. While there is rapid progress in natural language processing technologies that enable such interfaces, research in understanding writer interaction with such technologies and their effects is still emerging.

Recent studies have shown that writing and communication are influenced by the presence and content of suggestion systems \cite{arnold2018sentiment} — in terms of vocabulary use \cite{arnold2020predictive}, overall perceived sentiment \cite{arnold2018sentiment}, and interpersonal perception \cite{mieczkowski2021ai}. The focus of these studies has been the writing product. While there has been recent research on how the writing process is affected by a suggestion system, it has primarily been done through a behavioural lens, hence not revealing the decisions writers make that lead to the observed behaviour \cite{buschek2021impact}.
Writing research has long studied the writing process through a cognitive lens. The \citet{doi:10.1177/0741088312451260} Cognitive Process Writing Model (2012), a holistic cognitive model of writing, describes writing as a complex, non-sequential interaction between various cognitive writing processes.
In this study, we used the \citet{doi:10.1177/0741088312451260} model as an analytical lens to qualitatively understand how writers interact with an in-line next-phrase suggestion system, and compared writing with and without suggestions. Additionally, we studied how misalignment between sentiment bias in the model and the writer’s opinion affects writer-suggestion interaction.
\begin{figure*}
  \centering
  \includegraphics[width=11.95cm]{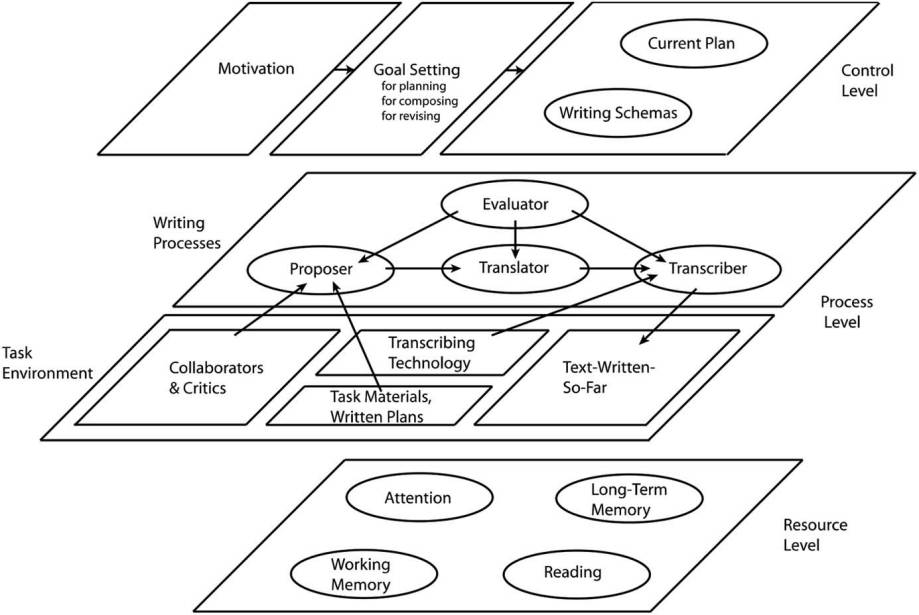}
  \caption{Cognitive Process Model of Writing by Hayes (2012)}
    \label{fig:Hayes Model}
\end{figure*}

We chose a movie review writing task and a suggestion system based on GTP-2 fine-tuned on the IMDb movie reviews corpus \cite{imdb}. We ask participants to watch two movies and then write a review for each, once with suggestions enabled and once without. We encouraged writers to express their personal opinions through their reviews and think aloud as they did so. Following that, we elicited retrospective protocols and analysed them with the think-aloud protocols through qualitative coding using a grounded approach. We found that writers interact with next-phrase suggestions in various complex ways: Writers abstracted and extracted multiple parts of the suggestions and incorporated them within their writing, even when they disagreed with the suggestion as a whole; along with evaluating the suggestions on various criteria. The suggestion system also had various effects on the writing process, such as altering the writer's usual writing plans, leading to higher levels of distraction etc. in addition to misalignment specific effects. Based on our observations, we propose a model of writer-suggestion interaction in the context of our study. Finally, we discuss how suggestion systems can be framed and analysed through the lens of our model, followed by a discussion of possible directions for research and opportunities for the design of suggestion systems.

\section{Related Work}
\subsection{Writing Research}
Research shows that writing is a complex, non-sequential process, with writers involved in different cognitive processes such as generating ideas, evaluating their ideas, converting ideas to language, etc., throughout a writing session. 

In a review paper, \citet{DonahueLillis+2014+55+78} outline four types of models that have been the most influential. These include text-oriented, social practice, didactic, and socio-cognitive models of writing. Donahue and Lillis suggest that while these models have a common goal of describing writing, their ‘empirical objects of study’ differ.

Among these, the \citet{doi:10.1177/0741088312451260} model (2012) is the cognitive end of the modern socio-cognitive spectrum and tries to describe the cognitive processes an individual writer engages in during the process of writing. Our paper aims to understand how individual writers interact with next-phrase suggestion systems. For this, our empirical object of study is the individual writer, and we view writing as a cognitive activity the writer engages in. To that end, the \citet{doi:10.1177/0741088312451260} Cognitive Process Model of Writing (CPMW) acts as a suitable analytical lens for our research. We give a brief description of the model below. For a detailed description, refer to \citet{hayes2015can}

\subsubsection{The Cognitive Process Model of Writing}
\emph{The \citet{doi:10.1177/0741088312451260} Cognitive Process Model of Writing (2012)} consists of 3 levels—the control, the process, and the resource level (Fig. \ref{fig:Hayes Model}). The control level includes components that direct or govern the activity of writing. The process level includes core writing processes and factors from the environment that influence the internal processes. The resource level includes crucial components for various human activities, including writing.
\\[0.07in]
\emph{Control Level: }
The control level contains Motivation, Goal Setting, the current writing plan, and the writer’s writing schema. \emph{Motivation} governs how engaged the writer would be in the process of writing. The writer decides the \emph{goal of the text} before composing (eg. an essay, thesis report) or during writing (planning, revision, composing). The goal of the text may also be assigned to the writer (e.g. write a movie review in our study). Writers often also come up with \emph{plans} for their text after they have set their initial goals (“I will start with an introduction and then make arguments”), which evolve and change through the process of composition. Depending upon its complexity, the writer may store this plan in their memory or write it down. \emph{Writing schemas} represent the writer’s knowledge of writing and composition. The schema informs the writer ‘how to write’ and attributes the composed text 'should' have—like writing strategy, genre, length, format, etc. (“The review \emph{should} start with a brief introduction”).
\\[0.07in]
\emph{Process Level: }
The process level consists of two sections: the \emph{internal writing processes} and the \emph{external task environment}. The \emph{internal writing processes} constitute the \emph{Proposer} which generates new ideas, the \emph{Translator} which converts the non-verbal, abstract ideas into coherent language, the \emph{Transcriber} which transcribes the generated language while following the rules of the written form (drawing letters, capitalisation, etc.) using different transcription tools (pen, keyboard etc), and the \emph{Evaluator} continuously evaluates the outputs of all the above processes and decides whether they are adequate and appropriate.

The \emph{external task environment} includes the writer's social and physical surroundings. For example, the comments and suggestions that collaborators or critics might give are part of the social task environment. Similarly, transcribing technology, reference material, written plans, text written so far, etc., become the physical task environment. In this paper, we are interested in studying the writer’s interaction with their task environment—i.e. the text editor enabled with next-phrase text suggestions.
\\[0.07in]
\emph{Resource Level: }
The resource level includes components representing attention, long-term memory, working memory, and reading, which are more general-purpose mental resources. \emph{Attention} represents our ability to maintain focus, often referred to as executive function. \emph{Long term memory} includes the writer’s repertoire of knowledge, including schemas, vocabulary, grammar, spelling, and facts. \emph{Working memory} is temporary storage to handle information needed to complete tasks. The proposer, translator, transcriber, and evaluator use the working memory.

Additionally, we define a \emph{Working Memory State (WMS)} as an addition to the \citet{doi:10.1177/0741088312451260} model in the context of writing with text suggestions. It represents the current composition state in the writer’s working memory when they encounter a suggestion. It may contain a partial or a complete proposal(idea) and a partial or a complete translation(language construction) of the text the writer was about to write when they encounter a suggestion. The WMS helps us better describe the interactions of the cognitive processes of writing and their outputs with the suggestions.
\subsection{Writer-Suggestion Interaction}

\subsubsection{Interactions and effects of suggestion systems on writing}
There is a growing body of research on how suggestion systems affect writing---with some studies looking at them as writing tools while others framing them as AI-mediated communication technologies. Recent work in this area has shown that writing with suggestion systems can lead to more predictable language and shorter sentences in the written product \cite{arnold2020predictive}; along with showing that a positive sentiment bias in the suggestion model leads to more positive writing \cite{arnold2018sentiment}. Recent work demonstrates the potential of AI-Mediated Communication to increase communication efficiency and use of positive language~\cite{hohenstein2021artificial} along with undermining some dimensions of interpersonal perception~\cite{hancock2020ai}.

In adddition to studying effects, there has also been a recent interest in studying writer-suggestion interaction. \citet{buschek2021impact} study interaction logs of writers writing emails with multiple parallel suggestion and reveal nine behaviour patterns of interaction with suggestions. Some of aforementioned studies have provided preliminary evidence of writers using suggestions as prompts for ideas and language~\cite{buschek2021impact, arnold2016suggesting}. Recent work by \citet{singhglassman} observes writers taking various forms of 'integrative leaps' for incorporating suggestions into their writing during creative story writing with multi-modal suggestion systems(text, image and audio). We aim to add this growing body of work by studying writer-suggestion interaction from a 'cognitive processes' lens. This motivates our first research question: \textbf{\emph{How do writers interact with inline next-phrase suggestions, and what governs these interactions?}}

In addition to understanding how writers interact with suggestions, we are also interested in understanding how the presence and the content of suggestions affect the writing process, compared to a writing process without suggestions. Bringing us to our second research question: \textbf{\emph{How do suggestions and the subsequent writer-suggestion interactions affect the writing process with suggestions as compared to the writing process without suggestions?}}

\subsubsection{Studying writer-suggestion mis/alignment}
Text suggestions based on language models have been known to inherit biases and dominant ‘views’ from the text corpus used to train the language model \cite{basta2019evaluating, kurita2019measuring, sheng2019woman, hutchinson2020social}. Recent HCI literature has called for studying not just how the presence but the content of the suggestions affects people who use them~\cite{arnold2018sentiment}. Controlling the content of the produced suggestions by controlling the bias for comparing how a change may affect the writing process is therefore essential to our study. That said, a bias is inherently normative---a writer may deem a suggestion generated by a language model as good or bad based on their normative framework, or social values \cite{blodgett2020language}. Our interest, therefore, lies in the alignment (or lack thereof) between the writer and suggestion system and how such misalignment affects writer-suggestion interaction. This brings us to our final research question: \textbf{\emph{How does the degree of misalignment between writers opinion and model's bias affect writer/suggestion system interactions?}}
\subsubsection{Methodological choices in studying writer-suggestion interaction}
Recent approaches of studying writer interaction with language models using interaction logs \cite{buschek2021impact} and tools for doing so \cite{lee2022coauthor} consider interaction logs and writing session replays as ground truth for interpretive analyses. Having its roots in approaches like progression analysis~\cite{perrin2003progression}, these approaches differ from traditional HCI approaches such as contextual inquiry, where writers are interviewed after the writing task \cite{calderwood2020novelists,clark2018creative}. While these approaches provide an objective representation of writer-suggestion interactions at a behavioural level, they lack insight into the inner cognitive decision-making processes that writers might engage in.  

Qualitative writing research has used concurrent, and retrospective protocol analysis—first introduced by Ericsson and Simon \cite{ericsson1992protocol}— as a method for understanding the decisions made by writers and getting a peek into the writer’s cognitive processes. While the concurrent think-aloud method elicits more protocol segments and insight into the steps leading to the final decision, retrospective protocols provide more insight into the final decision  \cite{10.2307/1423365}. While research has shown that writing while thinking aloud doesn't necessarily affect the written product, prompting writers to verbalise \textit{too often} can result in the writing process being slower than usual \cite{ransdell1995generating}. A researcher thus has to find a balance between eliciting enough protocol statements while not letting the writer deviate from their usual writing process. (We describe how we arrived at this balance through our pilot in the methods section)

\subsection{Working with Indian non-native English speakers}
While studying writing tasks, research often categorises writers and speakers of English as either L1 speakers who are also often native speakers (typically people from the US, UK, or Australia)or L2 speakers, whose country of origin is typically a non-English speaking country (like China, Japan or Germany) \cite{ZHAO201747}. 

Work by \citet{buschek2021impact} demonstrates that L2 speakers accept and use more suggestions than L1 speakers, along with finding them more helpful than their L1 counterparts. Past research comparing the writing process of L1 and L2 writer shows L2 writers pausing after shorter writing episodes and lengths of writing continuous text as compared to L1 writers~\cite{chenoweth2001fluency, leijten2019analysing}. This might lead to L2 writers having a higher probability of encountering suggestions. Past research shows L2 writers using reference materials as language repositories\cite{plakans2012close}, where they may note down potentially useful phrases and words from the text for later use \cite{flowerdew2007language}, along with behaviours such as ‘patchwriting’, characterised by non-malicious or unintentional appropriation of text without explicitly citing it. L2 writers could also use suggestions as reference materials, potentially finding them more useful than L1 writers.

Our study is conducted in India, most of our participants are L2 speakers. Given the aforementioned background on L2 writers, we are motivated to answer our research questions in the context of non-native Indian English speakers, numbering 129 million according to the 2011 census \cite{wikipedia_2021}.

However, it is important to mention that non-native English speakers in India may differ from non-native speakers in other non-English-speaking countries. India is a region of immense diversity in language. Hence, English remains an official language for communication between states and is the language of choice for administrative services, law, and education. While all of our participants had a first language (or mother tongue) that was not English (e.g. Marathi, Hindi, Assamese, etc.), they had completed their K-12 schooling with English as the medium of instruction. We believe this differentiates Indian L2 English speakers and writers from L2 writers from other non-English speaking countries, and the findings should be interpreted as such.
\\[0.07in]
In summary, we intend to study three things.
\begin{enumerate}
    \item{Indian non-native English writers' interaction with next-phrase suggestions}
    \item{The qualitative differences between them writing with and without suggestions}
    \item{The effect of misalignment between the writer and the suggestion system on the writer-suggestion interactions}
\end{enumerate}
Our research aims not to test or validate particular hypotheses or approaches for giving text suggestions but to collect systematic observations and construct knowledge inductively on the writer’s interactions with text suggestions using a grounded approach. We present observations that can act as a theoretical springboard for further empirical research.

\section{Method}
To investigate our research questions, we developed two instances of suggestion systems --- one instance fine-tuned on positive movie reviews and another on negative movie reviews. We asked participants to watch two movies and write two movie reviews — one review without suggestions and another with. We asked writers to rate the movies they watched before writing the review and allotted the suggestion system such that we got a distribution of writers along an ‘alignment spectrum’. We collected concurrent think-aloud protocols while the users wrote their reviews and retrospective think-aloud protocols after they had finished writing. We qualitatively analysed these with the screen recordings of the writers’ writing process. We use the \citet{doi:10.1177/0741088312451260} Cognitive Process Model of writing (2012) as a theoretical framework to analyse our qualitative data. Below, we provide the details of this method.

Our primary goal was to create opportunities for diverse writer-suggestion interaction to capture how writers interacted with the suggestions and their content. We also wanted to observe instances where the writers had varying levels of misalignment with the suggestion system  — to see if and how writers used these suggestions when they agreed with them and when they did not. We chose a movie review writing task for the same. A movie review task is appropriate for many reasons. Firstly, writing movie reviews involves expressing one's opinion and arguing for it. While movie review writing may not be a usual task for most writers, it is analogous to other opinion expression and argumentation tasks that writers may engage in. Moreover, unlike other writing tasks such as creative story writing \cite{clark2018creative}, movie reviews have a star rating associated with them, giving us a linear scale of sentiment that may be used to create a variety of writer-suggestion misalignments. Secondly, countless movie reviews exist on the internet, along with available datasets \cite{maas}, giving us a rich resource to train a language model. Lastly, a movie watching and reviewing task is simple to assign and execute for participants.

We recruited amateur writers who agreed to write two movie reviews and think aloud as they did so. All writers were asked to write their first review without suggestions and the second with suggestions. In the with-suggestions condition, the writers were randomly assigned to a suggestion system trained on a corpus of movie reviews with a mean rating of either 2.5 or 8.5 (the Corpus Rating) \cite{10.2307/1423365}. The Corpus Rating is a representative number calculated by averaging the ratings in the positive or negative review corpus. Writers also rated each movie. Thus, this was a mixed-design qualitative experiment. The within-subjects variable was the presence or absence of suggestions in the editor. The between-subjects variable was the degree of misalignment between the writer’s rating and the mean rating of the corpus, defined as the following. We used the Corpus Rating to calculate the misalignment score, which represents how misaligned the writer’s opinion is with the bias in the suggestion system. Here, \emph{DoM: degree of misalignment, WR: Writer Rating and Corpus Rating: Corpus Rating}
\[DoM = WR - Corpus Rating\]
A degree of misalignment close to 0 indicates that the writer’s opinion aligned with the mean rating of the corpus on which their suggestion system was trained. A negative value tells us that the writer’s opinion about that movie was worse than the mean rating, and a positive value tells us that the writer’s opinion was better.
\begin{figure*}
  \centering
  \includegraphics[width=\linewidth]{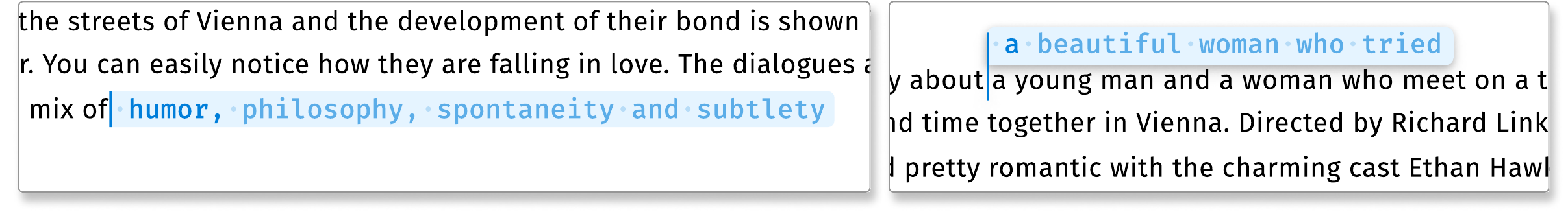}
  \caption{The interface presented to participants, with end-of-text suggestions (left) and middle-of-text suggestions (right)}
	\label{fig:apparatus_interface}
\end{figure*}
After the writer filled out an informed consent form, they were sent a link to their first movie. The writer was requested not to read any online reviews or view online ratings of the movie. We gave the writer about 24 hours to watch the movie. After the writer had watched the movie, we asked them to fill out a form where they had to rate the movie on a scale of 1 to 10, where 1 represents that the writer did not like the movie at all, and 10 represents that they liked it very much. The writer was additionally asked to choose one among the following options to ensure that there was no misunderstanding about the scale: (1) I really liked the movie, (2) The movie was average, and (3) I did not like the movie.

Next, we fixed a time with them for a review writing session. We conducted the two sessions over Zoom calls. In the first session, we sent the link to the assigned online text editor. The writer was asked to share their screen. The sessions were recorded for analysis. In the with-suggestions condition, we first demonstrated the suggestions interface (described below). The writer was then asked to practise writing a few sentences describing their day as a familiarisation task. After that, the writer was asked to write the movie review and think aloud as they did this.

After the first session, we shared the video recording of the session (without audio) with the writer and asked them to explain their writing process retrospectively. If the writer forgot what they had thought or if they contradicted themselves from what they said earlier, audio from that portion of the recording was played to trigger relevant memories.

We collected concurrent and retrospective think-aloud protocols as our primary source of qualitative data. The concurrent protocols, retrospective protocols, and video recordings helped us triangulate the data to get a clear picture of the actions and decisions the writers were making and why they were doing so. During a pilot, we realised that writers often forgot to think aloud and needed nudges. This resulted in writers explaining themselves in the middle of their writing process, thus interrupting their usual writing process. To counter this problem, we reframed the ‘think-aloud’. We instead asked the writers to ‘talk to themselves’ while writing along with explicitly asking them not to explain their decisions to us. This resulted in suitable protocols while minimising distraction from the writing process.
The Institutional Review Board of our university approved our study.

\subsection{Movie selection}
We chose 30 movies for our study as an initial set. We chose movies that were not widely released and ‘mainstream’, to ensure that the writers had not watched the movie previously and their reviews were not affected. We also avoided polarising genres such as historical dramas, war, political movies, animated movies, and documentaries. We picked movies from those released between 1990 and 2021. We rated the chosen movies based on ratings from three established sources: IMDb \cite{imdb}, Rotten Tomatoes (Critic and Audience scores) \cite{rottentomatoes}, and Letterboxd \cite{letterboxd}. Ratings from all three sources were averaged on a scale of 1 to 10 and utilised for final rankings. Based on this rating, the movies were divided into three categories: \textbf{Bad} (1-3), \textbf{Neutral} (4-7), and \textbf{Good }(7-10).We randomly chose three movies from each category, resulting in the nine movies listed in Table \ref{tab:movie_set}.
\begin{table}
  \caption{The selected set of movies and their aggregate internet ratings}
  \label{tab:movie_set}
  \begin{tabular}{lc}
	\toprule
	Movie Name&Aggregate Internet Rating\\
	\midrule
	Loqueesha \cite{imdb_2019} & 1.8\\
	Future World \cite{imdb_2018_5} & 2.5\\
	Glitter \cite{imdb_2001} & 3.2 \\
	Literally, Right Before Aaron \cite{imdb_2018_6} & 4.3 \\
	The Last Shift \cite{imdb_2020} & 5.5\\
	Nothing \cite{imdb_2004} & 6.1\\
	Little Miss Sunshine \cite{imdb_2006} & 8.1\\
	Hunt for the Wilderpeople \cite{imdb_2016} & 8.1\\
	Secrets \& Lies \cite{imdb_1997} & 8.4 \\
  \bottomrule
\end{tabular}
\end{table}
\subsection{Apparatus}
Our interface consists of a text editor capable of providing phrase and word completion suggestions both at the end of and in between text (Fig. \ref{fig:apparatus_interface}). We built three versions of the text editor, one with suggestions powered by a language model trained on an IMDB review corpus \cite{maas} with reviews with an average rating of 2.5, the second with an average rating of 8.5, and the third without suggestions. The interface provided word complete as well as next-phrase suggestion capabilities. The suggestion pops up as highlighted text near the top of the cursor when between text and in-line at the end. In our initial pilots, we deployed a suggestion interface where the writer had to press tab to select the whole suggestion.
\begin{table*}
  \caption{Comparing the positive and negative alignments of the system to the base GPT-2 suggestions}
  \label{tab:model_suggestions}
  \begin{tabular}{p{4cm}p{9cm}}
	\toprule
   \parbox{4cm}{\raggedright{\textit{Sample from GPT-2 base model}}} & \parbox{9cm}{I've been doing a lot of work on this blog over the last few years. One of the things that I've been working on is making sure that all of the posts that I make on this blog are written by people who have had the pleasure of writing for me in the past.}\\
	\midrule
	\parbox{4cm}{\raggedright{\textit{Sample from IMDb fine-tuned model (\textcolor{blue}{Corpus Rating = 8.5)}}}} & \parbox{9cm}{I saw this movie at the Tribeca Film Festival, and it was one of the \textcolor{blue}{funniest} movies I've ever seen. The acting was \textcolor{blue}{good}, and the story line was \textcolor{blue}{funny}. It was a lot of \textcolor{blue}{fun}, and I \textcolor{blue}{recommend} it to anyone who likes comedy.}\\
	\midrule
   \parbox{4cm}{\raggedright{\textit{Sample from IMDb fine-tuned model \textcolor{red}{(Corpus Rating = 2.5)}}}} & \parbox{9cm}{I \textcolor{red}{don't know where to begin} with this movie, it's a \textcolor{red}{complete waste} of time and money. I \textcolor{red}{don't know how anyone could make a movie like this.}}\\
	\bottomrule
\end{tabular}
\end{table*}
However, we received feedback that writers often wanted to select only the first few words in the sentence and had to delete the last few words after they ‘tabbed’ the suggestion. To resolve this issue, we designed the interface to select only a single word from the phrase when a writer pressed tab. To visually convey this interface behaviour, we highlighted the first word and added interpuncts (·) between words to represent this stepped approach to accepting a suggestion. (Fig. \ref{fig:apparatus_interface}). In our pilot, we observed that when users were typing continuously, the time between two keystrokes --- even for slow typists from our pool of participants --- was less than 300ms; i.e. an inter-keystroke time greater than 300ms usually meant a pause in typing. We wanted to generate a suggestion \textit{every time the user paused} in order to maximise user's encounters with suggestions. However, generating suggestions when users were typing continuously would not be of much use, as by the time a suggestion would appear, the user would've typed a new character. Hence, we decided to wait 300ms before a new suggestion was generated and displayed after a keystroke.

The suggestion system takes in the last 50 words the user has written to compute the suggestion using the language model running in the back-end. We make almost all computations on the server side to maintain low latency. We have used a GPT-2 \cite{radford2019language} transformer model for text prediction. To bias the model for the specific task of writing movie reviews, we used a fine-tuned GPT-2 model \cite{finetuned} on an IMDb movie review corpus with polar ratings \cite{maas}. The IMDb movie reviews dataset consists of 25,000 examples each of positive and negative movie reviews. These reviews are polarised; we selected positive reviews that contain reviews with a star rating greater than 6 (and averaging 8.5), and the negative reviews that contain reviews with star ratings less than 5 (and averaging 2.5). We fine-tuned each model for three training epochs and obtained a test perplexity score of 36.9713 for the positive model and 34.6978  for the negative model. We used the Hugging Face transformers library to train these models \cite{transformers}. We used the beam search algorithm to generate the most likely next phrase. The motivation to use beam search comes from the fact that we need to generate a few word completions. While methods like Nucleus Sampling \cite{nucleus} produce more varied text, these methods are more relevant to tasks of arbitrating text generation where the goal is to produce paragraphs worth of human-like text. In these cases, beam search repeats its predictions over and over after a few initial words, and therefore using sampling methods creates more human-like text. However, since our goal is to find the most optimal next phrase, given a previous prompt, we opt to choose a sampled version of beam search, which generates multiple potential candidates based on the top-5 beam scores and then samples from these candidates. Furthermore, modern systems like \cite{chen2019gmail} use beam search to generate completions. Using beam search in our experiment aligns our system with existing mass-deployed next-phrase suggestion systems, thereby adding external validity.

The confidence of beam search dictates the length of the suggestion. A beam search decoder maintains the probability score generated by the words added to the suggestion on every iteration. We empirically validate a threshold value, and new words stop being added when the probability score falls below this threshold value. We tuned the threshold value in a series of pilot tests by observing the Threshold value that does not create repetitions and aberrant words in the generations.

To verify how well the model inherits the bias, we performed a sentiment validation using a pre-trained BERT Sentiment Classifier. We provided prompts from the test set, which contained five hundred samples of reviews from both the positive and negative sections of the set and used our fine-tuned models to generate phrase completions for these prompts. We then passed these completions through a BERT-Large model trained for sentiment classification on the IMDB dataset. The reported accuracy of this pre-trained classifier on the IMDB validation set is 90\%. We observed that 76\% of the generations from the positively biased model were classified as positive, while 78\% of the negatively biased model were classified as negative. The remaining generations in both cases, classified into the category opposite to their generator model, can be attributed to the fact that the model tries to maintain coherence. In some cases, it is impossible to generate an oppositely biased candidate. Sample text generations are shown below (Table \ref{tab:model_suggestions}). These examples demonstrate the bias embedded in the respective language models. For examples from the study, refer to the findings section.

To verify the usefulness of our model in real-time phrase completions, we follow a metric similar to the one used by \cite{chen2019gmail}. We use our model on a set of pilot study data where users were asked to write reviews without a completion model. For each word that the user writes, we compute the next phrase. For every consecutive word of the phrase completion that matches what the user actually wrote, we increment the usefulness count by 1. We then add the user’s second word to the model input and repeat this process for subsequent words. We finally divide this count by the total number of words the user writes. Furthermore, we average this for five iterations of running phrase completion for each review. This score establishes how well the model can predict precisely what the user would have wanted to write. We obtain an average score of 0.3 for the positively biased GPT-2 model and a score of 0.29 for the negatively biased GPT-2 model. For a baseline comparison, the AWD-LSTM \cite{merity2017regularizing} model, fine-tuned on the entire IMDB corpus \cite{maas}, gave us a score of only 0.23.

\begin{figure}
  \centering
  \includegraphics[width=\textwidth]{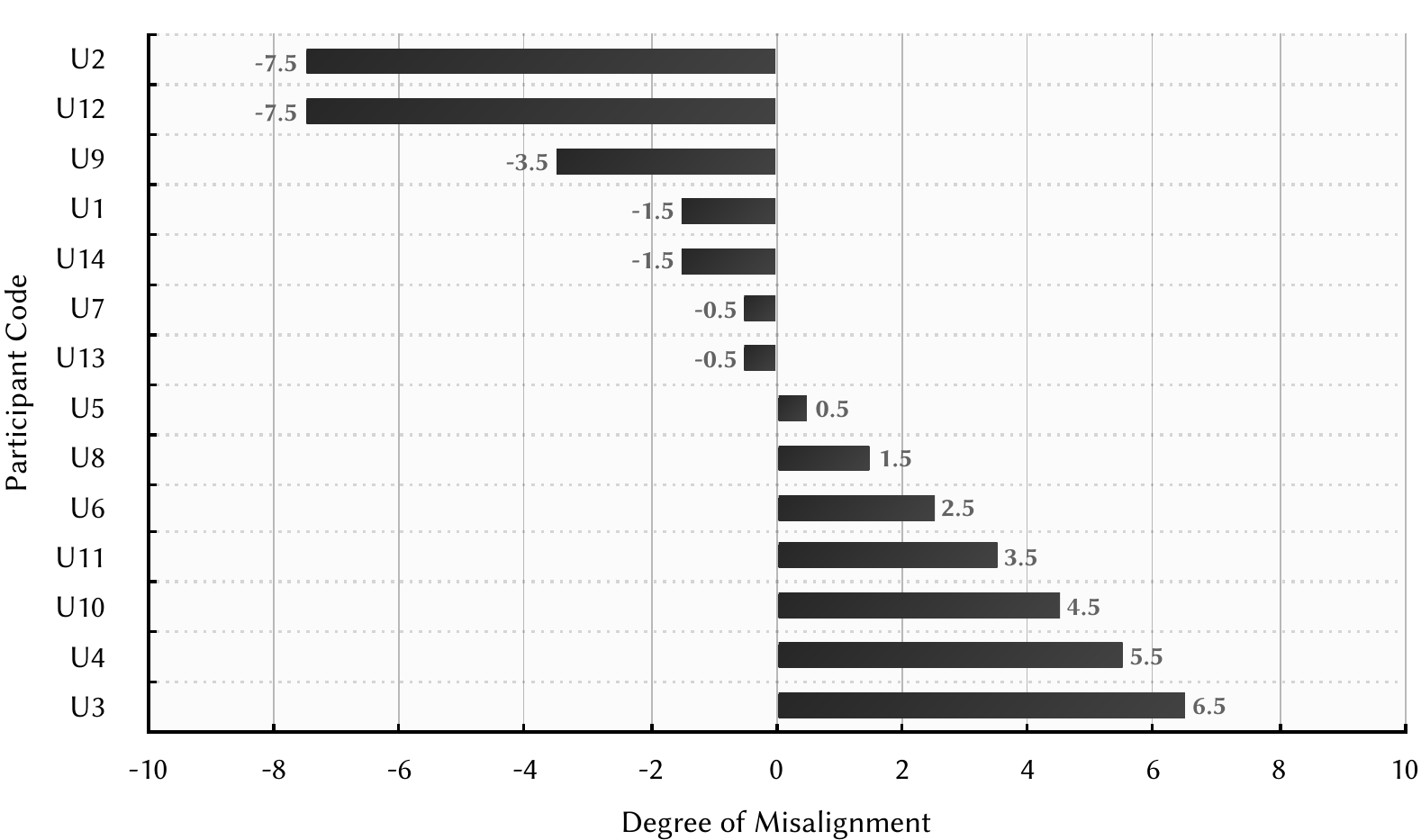}
  \caption{Distribution of participant degrees of alignment}
	\label{fig:chart}
\end{figure}

\begin{table*}
  \caption{Participant data on First Language, assigned movie, writer rating, system alignment, and the resulting degree of misalignment.}
  \label{tab:participant_data}
  \begin{tabular}{p{1.5cm}p{1.3cm}p{3cm}p{1cm}p{2cm}p{3cm}}
	\toprule
   \parbox{1.5cm}{\raggedright Participant Code} & \parbox{1.3cm}{Mother tongue}& \parbox{3cm}{\raggedright Assigned movie (for with suggestions review)} & \parbox{1cm}{\raggedleft Writer Rating (WR)} & \parbox{2 cm}{\raggedleft Corpus Rating for the system (Corpus Rating)} & \parbox{3cm}{\raggedleft degree of misalignment (DoM = WR - Corpus Rating)}\\
   \midrule
   \parbox{1.5cm}{\raggedright U1} & \parbox{1.3cm}{Marathi}& \parbox{3cm}{\raggedright Loqueesha} & \parbox{1cm}{\raggedleft 7} & \parbox{2 cm}{\raggedleft 8.5} & \parbox{3cm}{\raggedleft -1.5}\\
   \parbox{1.5cm}{\raggedright U2} & \parbox{1.3cm}{Marathi}& \parbox{3cm}{\raggedright Future World} & \parbox{1cm}{\raggedleft 1} & \parbox{2 cm}{\raggedleft 8.5} & \parbox{3cm}{\raggedleft -7.5}\\
   \parbox{1.5cm}{\raggedright U3} & \parbox{1.3cm}{Hindi}& \parbox{3cm}{\raggedright Secrets \& Lies} & \parbox{1cm}{\raggedleft 9} & \parbox{2 cm}{\raggedleft 2.5} & \parbox{3cm}{\raggedleft 6.5}\\
   \parbox{1.5cm}{\raggedright U4} & \parbox{1.3cm}{Hindi}& \parbox{3cm}{\raggedright The Last Shift} & \parbox{1cm}{\raggedleft 8} & \parbox{2 cm}{\raggedleft 2.5} & \parbox{3cm}{\raggedleft 5.5}\\
	\parbox{1.5cm}{\raggedright U5} & \parbox{1.3cm}{Malyalam}& \parbox{3cm}{\raggedright Hunt for the Wilderpeople} & \parbox{1cm}{\raggedleft 9} & \parbox{2 cm}{\raggedleft 8.5} & \parbox{3cm}{\raggedleft 0.5}\\
	\parbox{1.5cm}{\raggedright U6} & \parbox{1.3cm}{Telugu}& \parbox{3cm}{\raggedright Glitter} & \parbox{1cm}{\raggedleft 5} & \parbox{2 cm}{\raggedleft 2.5} & \parbox{3cm}{\raggedleft 2.5}\\
	\parbox{1.5cm}{\raggedright U7} & \parbox{1.3cm}{Marathi}& \parbox{3cm}{\raggedright Secrets \& Lies} & \parbox{1cm}{\raggedleft 8} & \parbox{2 cm}{\raggedleft 8.5} & \parbox{3cm}{\raggedleft -0.5}\\
	\parbox{1.5cm}{\raggedright U8} & \parbox{1.3cm}{Assamese}& \parbox{3cm}{\raggedright Future World} & \parbox{1cm}{\raggedleft 4} & \parbox{2 cm}{\raggedleft 2.5} & \parbox{3cm}{\raggedleft 1.5}\\
	\parbox{1.5cm}{\raggedright U9} & \parbox{1.3cm}{Gujarati}& \parbox{3cm}{\raggedright Loqueesha} & \parbox{1cm}{\raggedleft 5} & \parbox{2 cm}{\raggedleft 8.5} & \parbox{3cm}{\raggedleft -3.5}\\
	\parbox{1.5cm}{\raggedright U10} & \parbox{1.3cm}{Gujarati}& \parbox{3cm}{\raggedright Glitter} & \parbox{1cm}{\raggedleft 7} & \parbox{2 cm}{\raggedleft 2.5} & \parbox{3cm}{\raggedleft 4.5}\\
	\parbox{1.5cm}{\raggedright U11} & \parbox{1.3cm}{Hindi}& \parbox{3cm}{\raggedright Future World} & \parbox{1cm}{\raggedleft 6} & \parbox{2 cm}{\raggedleft 2.5} & \parbox{3cm}{\raggedleft 3.5}\\
	\parbox{1.5cm}{\raggedright U12} & \parbox{1.3cm}{Hindi}& \parbox{3cm}{\raggedright Loqueesha} & \parbox{1cm}{\raggedleft 1} & \parbox{2 cm}{\raggedleft 8.5} & \parbox{3cm}{\raggedleft -7.5}\\
	\parbox{1.5cm}{\raggedright U13} & \parbox{1.3cm}{Marathi}& \parbox{3cm}{\raggedright Little Miss Sunshine} & \parbox{1cm}{\raggedleft 8} & \parbox{2 cm}{\raggedleft 8.5} & \parbox{3cm}{\raggedleft -0.5}\\
	\parbox{1.5cm}{\raggedright U14} & \parbox{1.3cm}{Hindi}& \parbox{3cm}{\raggedright Future World} & \parbox{1cm}{\raggedleft 1} & \parbox{2 cm}{\raggedleft 2.5} & \parbox{3cm}{\raggedleft -1.5}\\
	\bottomrule
\end{tabular}
\end{table*}

\subsection{Participants}
We recruited participants who were university students or recent graduates, who had at least ten years of education in English and frequently watched English movies. All participants were reasonably fluent desktop typists and identified themselves as non-native English speakers — whose mother tongues were Marathi, Hindi, Malayalam, Telugu, Assamese, and Gujarati. Their K-12 schooling, however, was completed with English as a medium of instruction. Table \ref{tab:participant_data} contains the coded names of the participants along with the movie they watched for the with-suggestion condition, the rating they gave for each movie, the mean rating of the corpus on which their suggestion system was trained in the “with-suggestions” condition and the degree of misalignment. Figure \ref{fig:chart} shows the distribution of our participants across the alignment spectrum.
\subsection{Data Analysis}
The study resulted in (14 users x 2 sessions = 28 sessions), each with concurrent and retrospective protocols. Four of the authors conducted coding and analysis of these protocols. Before coding, the recordings were transcribed automatically using OtterAi \cite{otter} and cleaned up manually. During coding, we referred to the recordings when necessary. We also wrote extensive memos to document observations and construct theories.
\begin{table*}
  \caption{Different writer-system alignments with examples of suggestions offered: Text styled 'bold' is the suggested text, while the regular text is written by the writer. For example,  U7 (WR: 8, DoM 0.5) wrote: \textit{the pacing of the movie is a little slow}; the biased language model ‘spun’ it positively and completed the sentence: \textit{at times, but it’s a very good movie.}}
  \label{tab:alignments}
  \begin{tabular}{p{2.8cm}p{4.8cm}p{4.8cm}}
	\toprule
   \parbox{3cm}{\raggedright System (Right) vs Writer (Down)} & \parbox{4.8cm}{Positive}& \parbox{4.8cm}{\raggedright Negative} \\
   \midrule
   \parbox{3cm}{\raggedright Positive} & \parbox{4.8cm}{U5 (WR:9, DoM:0.5): This movie has all the makings of \textbf{a classic}}& \parbox{4.8cm}{\raggedright U3 (WR:9, DoM:6.5): ..the characters were decent. \textbf{However, the movie was so bad that I..}} \\
   \midrule
   \parbox{3cm}{\raggedright Slightly Positive} & \parbox{4.8cm}{U7 (WR:8, DoM:-0.5): the pacing of this movie is a little slow \textbf{at times, but it’s a very good movie}}& \parbox{4.8cm}{\raggedright U10 (WR:7, DoM:4.5): The lead singer is good, \textbf{but the rest of the cast..}} \\
   \midrule
   \parbox{3cm}{\raggedright Slightly Negative} & \parbox{4.8cm}{U9 (WR:5, DoM:-4): The acting is blunt and not very well done, \textbf{but the writing is also pretty good..}}& \parbox{4.8cm}{\raggedright U6 (WR:5, DoM:2.5): ...made me feel attentive all the time and I would not prefer the same. \textbf{This movie is a waste of time and money.}} \\
   \midrule
   \parbox{3cm}{\raggedright Negative} & \parbox{4.8cm}{U12 (WR:1, DoM:-7.5): There wasn’t a single thing done right in \textbf{the film, but that doesn’t stop it from}}& \parbox{4.8cm}{\raggedright U14 (WR:1, DoM:-1.5): …directed by James Franco and is a terrible film. \textbf{The acting is awful and the characters…}} \\
	\bottomrule
\end{tabular}
\end{table*}
We followed a three-phase strategy for coding and analysis. In the first phase, we created a codebook. To do this, first, we inductively open-coded eight writing sessions (four with and four without suggestion) of four participants. We then compared the transcripts of each writer's with-suggestion session and without-suggestion session to polish our codes. This also told us how the writer’s suggestion process differed from their without-suggestion writing process, highlighted habits and practices that were inherent to that particular writer’s writing process and new practices and habits resulting from the suggestions. We also generated deductive codes based on Hayes 2012 \cite{doi:10.1177/0741088312451260}. Finally, we compared and merged these two sets of codes to create a unified codebook. Thus, at the end of the first phase of our analysis, we had a codebook consistent with existing literature and codes that we had found through our study, along with extensive memos. In the second phase, we coded 12 sessions using the codebook to validate or update the codes. We continued writing memos to create provisional theories and compare them against our previous memos. The third phase included validating our existing codes and memos through new evidence in the remaining eight sessions. After the first four users, the data collection and analysis was done iteratively. By the time we completed 14 users, we believed we had reached saturation along with having a diverse range of alignments and stopped further data collection.

\section{Findings}
We begin our findings by presenting a few examples of suggestions generated during the writing sessions of eight of our participants presented in Table \ref{tab:alignments} as eight different writer-suggestion alignments leading to uniquely different generations and interactions. We provide the text they had written just previously for context.

Next, we report how suggestions contributed to the various writing processes as defined by Hayes(2012)\cite{doi:10.1177/0741088312451260}. We then describe how writers evaluated the suggestions for incorporating them into their text and what factors governed this evaluation. Finally, we compare the writers' process with and without suggestions to report some interesting effects we observed.

\subsection{Contribution of suggestions to the processes of composition}
We found that next-phrase suggestions contributed to the processes of \textbf{proposer, translator, and transcriber}, as described in the Hayes Model(2012). They augmented any, all, or different combinations of the processes of writing. These ranged from suggestions exclusively contributing to only one process (e.g., transcribing) to suggestions contributing to all three processes for a given sentence (e.g., writer tabbing through a complete sentence for which they had not generated either a proposal or a translation). They either accepted the contributions at the moment or ‘stored’ them for later use --- in their mind or the text editor.

\subsubsection{Contributions to the proposer}
Writers often abstracted a topic/theme present in a suggestion and used it as an inspiration for new sentences. When writers did not have a proposal, instead of coming up with one from scratch, they would abstract a topic from the suggestion and use that as a prompt to generate a proposal for their upcoming sentence. For example, U9 (WR: 5, DoM: -3.5) was suggested \textit{“...but the writing is also pretty good."} by the system. They rejected the specific opinion in the suggestion but picked up the theme: 'writing of the film' and expressed their own opinion about it, “the writing of the film was weird”. U9 (WR: 5, DoM: -3.5) later remarked:\textit{ “I saw the term writing and [thought might as well write about it]”.}

Along with abstracting themes to generate completely new proposals from the suggestions, writers also incorporated themes and opinions abstracted from the suggestions into the proposals currently in the working memory state (WMS). Needless to say, they translated these new proposals in their own language, without using the sentence structure or vocabulary from the suggestion, at times being more elaborate and nuanced in their writing as compared to the suggested text. For example, U4 (WR: 8, DoM: 5.5) started a sentence with \textit{‘The protagonists...’}. The system then floated a suggestion: ‘were different from each other.’  The writer agreed with the suggestion saying \textit{“makes sense to me”}, but wanted to mention the protagonists’ names. Therefore, they continued writing the sentence \textit{‘The protagonists, Dave Stanley and Jevon Williams,...’} and then incorporated the theme of 'difference between the protagonists' as seen in the suggestion, in their own argument while \textbf{adding more detail} and \textbf{justification for their opinion}. They wrote \textit{‘...were very different from each other not just in skin colour but in every realm of life.’}, followed by an expansion of the argument. During the retrospective protocol session, we often paused and asked writers to remember what they were planning to write before the suggestion appeared, using their concurrent protocols as memory aids. U4 confirmed that they planned on giving a general description of the characters before the suggestion appeared without emphasizing their differences for describing them. (\textit{“Had that not appeared, I would have probably gone on and given a general description of the characters.”})

In some cases, suggestions triggered a completely unrelated train of thought that led to a new idea, leading to proposal that were unrelated to, yet an effect of the suggestion.(U3 (WR: 9, DoM: 6.5): \textit{“the suggestion said ‘it’s not funny, it’s not interesting’ and I think that ‘not funny’ part made me think of how the movie is not a comedy and it’s a drama”.})

While suggestions prompted new proposals when there was lack of one, along with augmenting an existing partial of complete proposal, we did not observe writers abandoning existing proposals in the WMS for themes or ideas from the suggestions. Unlike for the proposer, we did observe writers abandoning their original translations to prefer the translation from the suggestion.

\subsubsection{Contributions to the translator}

Besides abstracting themes and opinions from the suggestions, we also observed writers abstracting the sentence structures they observed in the suggestions and using them in their own writing. For example, in the case of U3 (WR: 9, DoM: 6.5), they received a suggestion: \textit{‘great acting, good storyline and a great cast’}. U3 abstracted this peculiar structure of 'comma-separated adjective-noun pairs' describing different features of the movie and wrote \textit{‘great acting, effective cinematography, beautiful music.'} Here, they did not necessarily pick up the opinion from the suggestion, nor did they write about the same topics --- they solely abstracted the presentation of the sentence.

Along with abstracting sentence structures, writers also extracted vocabulary and phrases from the suggestions and used them to express their ideas (i.e., translate their proposals.) This was evident when we compared writers' with-suggestion reviews with their without-suggestion reviews. We observed increased use of \textit{“classic movie-review language.”} (as described by one of the writers - U5 (WR: 9, DoM: 0.5)); such as \textit{"a must-see for all ..."} (U5), \textit{"... is top-notch."} (U7), "... I wouldn't really recommend it" (U9) etc. 

While many writers gave opinions on specific film-making elements in both their reviews, their sentences in the without-suggestion trials were longer, more nuanced, structurally complex, descriptive, personalised / localised in time, and with rationale. On the other hand, the suggestion system largely and frequently offered sentences that were single-noun-single-adjective  \textit{(“the acting was [...]”,  “the writing was [...]”, “the story was [...]”)}, and this language was adopted by most authors in the with-suggestion trials, with two instances of writers writing the exact same phrasing to express their ideas: \textit{“the acting was top-notch”} by U5 (WR: 9, DoM: 0.5) and U7 (WR: 8, DoM: -0.5). These observations corroborate findings by \citet{buschek2021impact} that corpus-specific wordings replaced participants’ own style of writing when they write with suggestions. Table \ref{tab:comparing} illustrates the contrast in language between with- and without-suggestion essays of the same writers with two examples.

\begin{table*}
  \caption{Comparing U8 and U5’s without and with-suggestion tasks, and how suggestions influenced language. Outputs in without-suggestion trials were longer, more nuanced and structurally complex, while their with-suggestion trial outputs were shorter, single-noun-single-adjective sentences.}
  \label{tab:comparing}
  \begin{tabular}{p{2cm}p{9cm}p{2cm}}
	\toprule
   \parbox{2cm}{\raggedright Participants} & \parbox{9cm}{Without Suggestions}& \parbox{2cm}{\raggedright With Suggestions} \\
   \midrule
   \parbox{2cm}{\raggedright U8} & \parbox{9cm}{\raggedright (WR: 7.5) The actors were phenomenal in their portrayal of the characters, it was truly brilliant acting and they were what really brought the film together. It really showed that the actors had a great understanding of their characters. All of them having such distinct personalities, this was another possible trainwreck in the sense of it becoming too much for the viewers, again was dealt with expertise. The dialogues really helped to fully realise these characters-it revealed their inner motivations, thoughts, insecurities.} & \parbox{2cm}{\raggedright (WR: 4, DoM: 1.5) The acting was flat and the writing was terrible.}\\
   \midrule
   \parbox{2cm}{\raggedright U5} & \parbox{9cm}{\raggedright (WR: 8) A nicely paced slice of life movie, it was a nice break amongst all the action, adventure, and sci-fi I've seen lately. It's deeply grounded in reality and offers commentary on things like racism, which is sort of shown from the eyes of both Jevon and Stanley. Even though it was a slow-ish drama, the pacing was nice, and I did not feel bored while watching it. And being a slice of life movie just added to its simplicity.}& \parbox{2cm}{\raggedright (WR: 9, DoM: 0.5) The acting is top notch and the story is simple yet engaging.} \\
	\bottomrule
\end{tabular}
\end{table*}

This also echoes with work by \citet{plakans2012close} who suggest L2 writers (which was the demographic of our study) use reference materials as language repositories; that is, they identify useful phrases from reference materials that can be used at the moment or for future use. As described earlier, we, too, observed writers exacting words and phrases from the suggested text. This suggests that L2 writers may also consider suggestions as language repositories for borrowing phrases from.


We found many instances of writers abandoning their original translation and choosing a translation provided by the suggestion, unlike proposals. Writers' reasons for doing so were: to save typing effort when suggestions \textit{"were more or less faithful"} to their proposal or better expressed their proposal than their own translation. For example, U5 (WR: 9, DoM: 0.5) explained how they selected a suggestion that conveyed their intentions better than their original translation: \textit{“I was writing ‘it was warm and funny’. [...] After writing ‘funny’, I noticed the word ‘heartwarming’ in the suggestions and I thought that was better. [So I selected] ‘and heartwarming’ [and later deleted] ‘warm and’)} thus changing their original translation from \textit{‘warm and funny’} to \textit{‘funny and heartwarming’.}

\subsubsection{Contributions to the transcriber}
At times, the suggestion or parts of the suggestion precisely matched the translation in the writer's WMS. In such cases, the writer pressed ‘tab’ to select the whole or parts of the suggestion to save typing effort. For example, U3 (WR: 9, DoM: 6.5), while writing about the plot, typed \textit{‘The movie follows the lives of a dysfunctional family that is…’} when they got the suggestion, \textit{‘...trying to figure out what’s going on’}. U3 tabbed through \textit{‘trying to figure out’} and then typed \textit{‘how to be happy together.’} U3 noted that part of the suggestion exactly fit the content and position of the translation in mind and used the suggestions to reduce typing effort. The writer did not have to completely agree with the proposal suggested by the system, but just bits of its translation, such as a few words, a single word, or part of a word.

Transcription-level contributions also occurred in the form of word completes. A writer would type out the first few letters of a word, and the suggestion would complete it. For example, U4 (WR: 8, DoM: 5.5) typed ‘per’, and the system completed the word ‘perception’. The suggestion exactly matched the writer’s intended translation; hence, the contribution was only on a transcription level.

\subsection{Evaluating suggestions}
Since the suggestions came from the system, writers hadn't proposed, translated, or transcribed the suggested text themselves and had to evaluate them on several criteria. They evaluated the suggestions independently or against the text they had written so far or compared them to their WMS. Their beliefs about the system also governed how they evaluated the suggestions. This section presents how writers evaluated the suggestions and the criteria they used.

\subsubsection{Independent evaluation of suggestions}
Writers did an “independent evaluation” of a suggestion when they encountered it at a time when they had no ready proposal or translation in their WMS. This independent evaluation involved no comparison to the text written so far and looked at each suggestion as an independent piece of text to be evaluated. Following were some of the interesting criteria for suggestion evaluation.
\\[0.07in]
\emph{Misalignment: }As expected, misalignments between the writer’s rating and the average corpus score increased the chances of writers rejecting suggestions. Writers could notice when the misalignment was huge. For example, U3 (WR: 9, DoM: 6.5) loved the movie (rating it 9) but was assigned a system with Corpus Rating 2.5 (degree of misalignment = 6.5). U3 remarked, \textit{“Suggestions were continuously telling me it was a very bad movie... I was like no, it was not a really bad movie”.} For other users, a misalignment in the intensity of the sentiment led to several rejected suggestions. U8 (WR: 4, DoM: 1.5) had given the movie a rating of 4 (degree of misalignment = 1.5). This meant they disliked the movie, but the suggestion system was more negative than their opinion. At several encounters with the suggestion, U8 exclaimed aloud \textit{“Wow!”}and rejected it. They explained in the retrospective narration,\textit{ “[the suggestion was] very blatant, and I didn't want to be this blatant”}. Thus, misalignment became an important criterion for evaluation. 
Besides the sentiment misalignment, writers also disagreed with specific opinions or certain details in the suggestions. For example, U3 (WR: 9, DoM: 6.5), while they were describing one of the locations from the movie, wrote ‘Its a place where’, and were suggested ‘black people go to meet other black people.’ To this, during their retrospective protocol, U3 exclaimed: \textit{“Why does it have to be specially ‘black people [go to meet] black people?' Why can't [it] be ‘people go to meet other people?’ [...] It's a place where two people as different as Dave and Jevon meet up.[...] I wanted to highlight that they are very different people. But their skin colour is not what is dividing them. It's actually the thing that connects them.”}
\\[0.07in]
\emph{Movie review schemas: }Writers had mental schemas about what content was appropriate to put in a movie review, whether they were writing with- or without-suggestions. Suggestions going against this schema were often rejected. For example, a common writing schema for our participants was to not give out spoilers in a movie review. When U3 (WR: 9, DoM: 6.5) got a suggestion saying \textit{‘For example’}, they rejected it, with the following explanation: \textit{“I was like I’ll maybe give an example but I am like no I wouldn’t want that in a movie review which I would read. I don’t want spoilers to that degree so I didn’t write”}.
\\\
Writers had a schema or a rough plan for the structure of their review. They utilised it to evaluate the position of the suggested text, independent of the text preceding it. For example, U5 (WR: 9, DoM: 0.5) was suggested \textit{‘this is one of the few movies that I have …’} at the beginning of the review, but they had a different plan. They rejected this suggestion saying \textit{“[I want to write] a two-line plot summary of the movie. [...] that's a good way to hook [the] audience in with a little bit of a story.”} However, as discussed above, even after rejecting a suggestion for their position, writers would ‘store the suggestion for later use’ — either in their memory or on the text editor.
\\[0.07in]
\emph{Factual accuracy and truth value: }Often, suggestions were rejected because they were factually inaccurate. This includes inaccurate details about the movie, wrong dates, and inaccurate references to the real world. For example, when suggested - ‘\textit{the special effects were…’,} U3 (WR: 9, DoM: 6.5) rejected the suggested text, saying \textit{“...because there were no special effects”.} In another instance, U6 (WR: 5, DoM: 2.5) rejected a suggestion - \textit{‘..waste of ticket money’,} and explained that one could not watch the movie in the theatres as the world is in the middle of a pandemic: \textit{“… it's 2020 [no movies are running in theatres right now]. So you won't really waste your money.”}

\subsubsection{Evaluation with respect to text written so far}
Writers also evaluated the suggestions for consistency and flow with respect to the text they had written so far. For example, when U5 (WR: 9, DoM: 0.5) described the movie's plot in their first paragraph, the system suggested - \textit{‘This is one of my favourite movies of all’.} Although U5 had liked the movie, they declined the suggestion stating, \textit{“That [suggestion] seemed too abrupt a change [compared to the sentence I just completed].”} Conversely, in other cases, writers accepted suggestions because they flowed well with the preceding text: \textit{“I felt this [suggestion] had a good flow with the previous [text]).”} (U5 (WR: 9, DoM: 0.5))

\subsubsection{Evaluation with respect to the Working Memory State (WMS)}
Writers often compared suggestions with the partial or complete composition in their working memory: (U1 (WR: 7, DoM: -1.5) \textit{"I used to stop and think can I use this suggestion and integrate [it] into what I am thinking or not".} Writers either checked whether the suggested text was 'more or less' compatible with their WMS and whether the suggestions provided an improvement over their WMS. For example, U5 (WR: 9, DoM: 0.5) got a suggestion \textit{‘... story is simple yet engaging’.} Here, they accepted the suggestion as the suggestion sufficiently expressed their proposal: \textit{“I was going to write something about the story being engaging in the next sentence. [...] and [the suggestion] was almost similar”.} In the retrospective protocol, they then stated this as a general principle they used for accepting suggestions: \textit{“If I was going to write something similar or exact to [the suggestion] then I would take the suggestion. [...] The basic meaning of the sentence should remain the same.”}

Suggestions also helped writer improve their WMS by providing alternatives, especially for their translations. For example, U5 (WR: 9, DoM: 0.5) remarked: \textit{So I thought, [the word] 'interaction' again, but that didn't feel right. Then [the system] suggested "adventure" and I thought it was better, [so I accepted it].}

Our findings echoed the distinction that \citet{stratman19875} make between ‘reading for comprehension’ vs ‘reading for evaluation’. Reading for comprehension, as defined by \citet{stratman19875}, is the attempt to construct a “satisfactory internal representation of the meaning of the text”. Reading for evaluation usually encompasses several goals besides comprehension, such as fixing spelling or grammatical errors, vocabulary, and revision in general.
We found the state of WMS paramount in deciding whether writers read for comprehension or evaluation. In cases where a proposal and translation both existed, writers were less likely to be affected by offered suggestions. (U8 (WR: 4, DoM: 1.5): "…this point was there in my mind. I was like, I don't want any influence. I want to type it out."). If the writer was in the process of forming a proposal/translation, reading suggestions would distract them, making them lose track of their thoughts. (U4 (WR: 8, DoM: 5.5): \textit{“I was trying to form the sentence, and then I looked up [at the screen]. That’s when the suggestion distracted me.”}) To counter this problem, some writers deliberately ignored the suggestions. (U8: \textit{“I wasn’t reading [the suggestions]...If I had read them my flow would have been broken.”}) On the other hand, if the writer did not have any proposal, they were more open, and in some cases, would even rely on the system for their next proposal and translation(see § Effects). U3 (WR: 9, DoM: 6.5): "\textit{Partly because [...] I didn't know what to write. I didn't know where to go after the first sentence and I was thinking, [...] what do I write, but then the suggestion showed up, and I'm like yeah, that option works. So then, I went with that because I didn't have anything else in mind… like any definitive thing in mind. That’s why I did it.}" 

\subsubsection{Evaluation influenced by the beliefs about the suggestion system}
Writers had at least two beliefs about how the system worked, influencing how they evaluated the suggestions and how much they trusted them. A few writers believed (wrongly) that the suggestions reflected what most people on the internet thought about that specific movie. This belief made the suggestions seem grounded in reality and trustworthy.For example, U8 (WR: 4, DoM: 1.5) reasoned that they got all these negative suggestions because they wrote the name of the movie at the top, so the system searched Google (which has access to several reviews of the movie) and 'knew' that it was a bad movie. Likewise, U6 (WR: 5, DoM: 2.5) echoed \textit{“I think it starts with a search for that keyword [in this case, the title of the film] and then suggests related sentences or something like that"}. This belief, that the reason suggestions were negative was because many people have talked negatively about the movie, may have influenced U6 into writing more negatively: \textit{“[...] the suggestions were obviously related to how bad the movie was. And when I started writing about the movie, I did not think it was that bad of a movie. And when I was about to complete it, I found the reason why it was bad.”}) These writers were more likely to believe the suggestions were true representations of other people’s opinions on that movie, and were thus more open to accepting them.

On the other hand, some users (partially correctly) believed that the system would generate suggestions based on their initial few sentences. For example, U5 (WR: 9, DoM: 0.5) decided to ignore all suggestions (except word-completes) for their first two sentences while they established context for the movie and the review so that the system would be able to come up with ‘more accurate’ suggestions. \textcolor{black}{(U5 (WR: 9, DoM: 0.5): \textit{I just started writing the review, right? So then it'll [give me] generic sentences. [...] Yeah, so my rationale was that the [system] probably doesn't have any context for writing about [movie reviews].} So I thought I'll ignore them for the first two lines where I'm talking about the movie. Because [the suggestions] won't make sense anyway)}

Mirroring the same expectation, U9 (WR: 5, DoM: -3.5) was enraged when the system kept giving negative suggestions, even after they had written positively about the movie. Because of this, they deemed the system incompetent, and said,\textit{ “The first, like this time it was fine, but later I was annoyed. Just read what I have written, [mate]!”} As U9 trusted the system less, it led to a decrease in their openness to suggestions. U5 and U9 believed that the suggestion system would adapt to their writing by either making the suggestion more specific or more in line with their sentiment found the suggestions given enough context. These writers’ openness to suggestions decreased over time as the suggestions did not meet their expectations.

\subsection{Effects on the writing process}
\subsubsection{The role of alignment}
Without a suggestion system, writers found no reason to deviate from their original translation and the sentiment it conveyed --- if they liked the movie, they wrote as much; if they hated it, they wrote as such. However, the existence of suggestions led to changes in criticality in expression for many writers. We found that these changes depended heavily on the degree of misalignment with the corpus on which the system was trained.
\\[0.07in]
\emph{Aligned:} In the case that the writer’s sentiment was exactly aligned to that of the system (with DoM roughly from -2 to 2), we observed that writers did not realise that the system was biased. During the debriefing, we told our participants about the system's bias (positive/negative). Most writers claimed that they did not notice a bias. (U5 (WR: 9, DoM: 0.5): \textit{“I didn't notice any bias”}). Only when told that the system had a bias did they realise in retrospect that the suggestions did seem to be too positive/negative. (U5 (WR: 9, DoM: 0.5): \textit{“Now that you say [the system was biased], I notice there was nothing negative about the movie when [the system] suggested.”}). U14 (WR: 1, DoM: -1.5) described their writing experience as \textit{“I felt like [the system] and I were having a conversation and both hating on the movie together”}. When explicitly asked if they noticed that most suggestions were negative, they replied, \textit{“I didn’t think [the system] was biased, I thought [it] was because the system knew [the movie] was bad”.} On the other hand, in cases where the alignment was not perfect, writers noticed that the system was biased. For example, U2 (WR: 1, DoM: -7.5), who was writing a critical review, mentioned noticing a bias: \textit{“Even if I was putting in negative views, it was coming out positively” }
\\[0.07in]
\emph{Slightly Misaligned:} As part of our distribution, a few writers were aligned with the system’s bias, but not quite as strongly (degree of misalignment roughly between -3.5 to -0.5 and 0.5 to 3.5). In these cases, it seems that the writers were persuaded to write more strongly than they originally intended to. U7 (WR: 8, DoM: -0.5), who liked the assigned movie and had a positively biased system, was initially critical of the film \textit{(‘the pacing of this movie is very slow [...] 2hrs 20 minutes is a bit much’}), despite liking it and giving it an 8-star rating. However, the text suggested by the system reminded them that they were being too negative. While thinking aloud, U7 said: \textit{“I’m being too much of a downer”}), and altered their writing after that to be less critical and more positive \textit{(‘the plotline is well fleshed-out, the story is very immersive and believable’).}

As mentioned above, U6 (WR: 5, DoM: 2.5) was one of the writers who believed that the system’s suggestions were “reflective of the internet’s opinion on the movie”. U6 began their review with almost no criticism, writing a weakly positive review (U6: \textit{The movie was good but [I initially thought] it was a horror movie}). While they did not particularly like the movie, they stated that \textit{“all movies are a type of art”}, and felt it would be “rude” to begin the review with criticism. Midway, however, their sentiments changed, and they began writing more negatively.\textit{ (“And when I started writing about the movie, I did not think it was that bad of a movie. And when I was about to complete it, I found the reason why it was bad.”)}. The constant negative suggestions convinced them to become more negative \textit{(“I was convinced”)}, and that if there had been no suggestions at all, \textit{“[the review] would have been more positive”}. These observations echo results reported by \citet{arnold2018sentiment}, where people who were going to write ‘mildly’ positive content ended up writing content that was ‘clearly’ positive.

But such misalignment did not affect everybody. U8 (WR: 4, DoM: 1.5), who also believed the system was getting its information from the internet and received substantially negative suggestions, felt that they themselves were \textit{“not qualified to be so critical”}, since they weren’t an expert in the field. U8 rejected many strongly negative suggestions such as \textit{“the movie is just boring”} and modified them to statements such as \textit{“the movie did get boring to watch”}. Thus U8 could retain control and did not get swayed by the misalignment.
\\[0.07in]
\emph{Completely Misaligned:} However, when the system and the writer were completely misaligned (DoM roughly greater than 3.5 and lesser than -3.5), the suggestions did not particularly affect any writer’s sentiment — in most cases, it caused frustration with the system. U3 (WR: 9, DoM: 6.5) responded verbally to the system \textit{“No, it is not a bad movie!”.} Most writers in this category dismissed such suggestions with a completely opposite sentiment. However,  some did partially accept the sentence and flipped the sentiment of the sentence. U2 (WR: 2, DoM: -7.5) was offered: \textit{“..made it all the more interesting”}. They accepted most of the suggestion, but changed the last word to \textit{“tiresome”.} U3 (WR: 9, DoM: 6.5) was almost about to accept the suggestion \textit{“this is one of the worst movies I have ever seen”} and flip the sentiment, but realised that would result in a more positive sentiment than they intended \textit{(“this is one of the best movies I’ve seen”)}, and settled on a toned down version to match their intent \textit{(“it is a pretty good film”)}.
While this was true for most writers, we found one interesting case where even suggestions with completely misaligned sentiment affected a writer’s strength of opinion. U9 (WR: 5, DoM: -3.5) began the review with fairly neutral sentiment \textit{(“The movie was okay-ish”)}. When the suggestion system repeatedly offered up positive suggestions, they ended up frustrated \textit{(“[..why doesn’t it read what I have written..]”)}, and when given \textit{(“this movie is very good”)}, they exclaimed \textit{(“NO! It is not a good movie!”)} and wrote down \textit{“this movie is not good...”} as a reaction to the suggestion, ending up with a decidedly negative sentence. This polarisation can be attributed to their reaction to the misalignment with the system.

\subsubsection{Changes in plan due to suggestions}
Suggestions often affected the writer's plans. For example, in the without-suggestion condition, U3 (WR: 9, DoM: 6.5) expected to start the review by introducing the plot in the first paragraph, followed by their opinion on the movie — according to their schema for the structure of a movie review. \textit{(U3: “you can't really begin a review by saying, "Oh this sucked" or like "This was good", so [I wanted to write] an introduction just to sum up the movie.”).} In their with-suggestion task, however, U3 was offered the suggestion “\textit{this is one of the worst movies I have ever seen”} right after their first clause, followed by “\textit{the acting is terrible”.} The writer evaluated based on ‘sentiment match’, disagreed, and wrote \textit{“This is a pretty good movie”}. They did not evaluate the suggestion for its position here, though they had explicitly stated that starting a movie review with opinions did not match their schema. U3 later realised that they had changed their intended structure for the review and mentioned, somewhat self-critically: “\textit{It is now [that] I'm doing the thing which I probably would have done in the second sentence, which is actually explaining the plot”}

Other times, the participants willingly accepted this change in plan for reasons like saving time and effort. U5 (WR: 9, DoM: 0.5), for example, accepted \textit{‘the acting was top-notch’,} because they felt it was \textit{“good enough to go with”}. When probed about the change in plan, they said that they \textit{“would have written about the acting in the next few sentences anyway”.} U5 reflected, \textit{“I would [normally] spend a little bit more time thinking about sentences, but here when it starts suggesting, I thought I would use those and try to incorporate the rest around it to save more time.” }

\subsubsection{Distraction due to suggestions}
When writing without suggestions, writers generated proposals and translations themselves, with minimal external influence on the proposer and the translator. The strategies they usually used to propose ideas were recounting movie details from their long-term memory (U7 (WR: 8, DoM: -0.5): \textit{I was kinda just mentally going through the entire movie}), creating a plan before writing (U3 (WR: 9, DoM: 6.5): \textit{Oh yeah, I'm thinking of where to begin. It's a lot of facets to cover)}, or reading the text written so far (U8 (WR: 4, DoM: 1.5): S\textit{o whenever there's a break in thought, I usually go back and re-read what I wrote)}. Similarly, writers used strategies like 'blurting' to capture ideas quickly and then iterating and revising (U8: “I just typed to keep my thoughts going. Because if I stop typing, then it feels like a proper block.”) for translating their ideas into language.

However, in the with-suggestions condition, constantly reading and evaluating suggestions led to higher cognitive load and distraction (U4 (WR: 8, DoM: 5.5): \textit{“I was trying to form the sentence. That's when I looked up. And I saw the suggestion, and I got distracted. To be honest with you, I was kind of judging this suggestion.”}). Suggestions also hindered the writers’ proposal generation process. (U8:\textit{ “I feel like if I'm thinking of something as I'm trying to form that thought, and I see something, which is completely different, I lose that thought which I was going with.”}) Especially when the content of the suggestions was unrelated to the proposal. (U8:\textit{“the suggestions are in a totally different direction and that kind of threw my flow off”}). To avoid this, U8 deliberately ignored the suggestions when coming up with proposals and translations. (U8:“\textit{I didn't want to compromise on the phrasing of it, I think, which is why I was like, kind of trying to ignore the suggestions.”}) The persistent and distracting nature of suggestions tended to hurry U8(WR: 4, DoM: 1.5) up. To avoid this, they would quickly develop translations and move on to the following sentence. They feared that the longer they took to form a sentence, the more chances there would be of accidentally reading and subsequently getting distracted by the suggestion. (U8: “\textit{The fact that I was getting suggestions [...]that made me want to finish the point quickly and go to the next point.”}) Due to this, it was difficult for a few writers to use an iterative approach to writing. (U6(WR: 5, DoM: 2.5): \textit{How I write is like, I write whatever I get into my mind and then correct it. [...] I had an issue while thinking because I've been reading all these suggestions, and I couldn't think straight because of the suggestions}).

\section{Discussion}
\subsection{A writer-suggestion interaction Model}
We found that writers interacted with inline next-phrase suggestions in various ways. They, directly and indirectly, used suggestions as aids for proposing, translating and transcribing by abstracting (themes, sentence structures etc.) and extracting (phrases, vocabulary etc.) from the suggestions. They also evaluated suggestions on several criteria before incorporating them into their writing. This use of suggestions affected their writing process in various unexpected ways.
\begin{figure*}
  \centering
  \includegraphics[width=\linewidth]{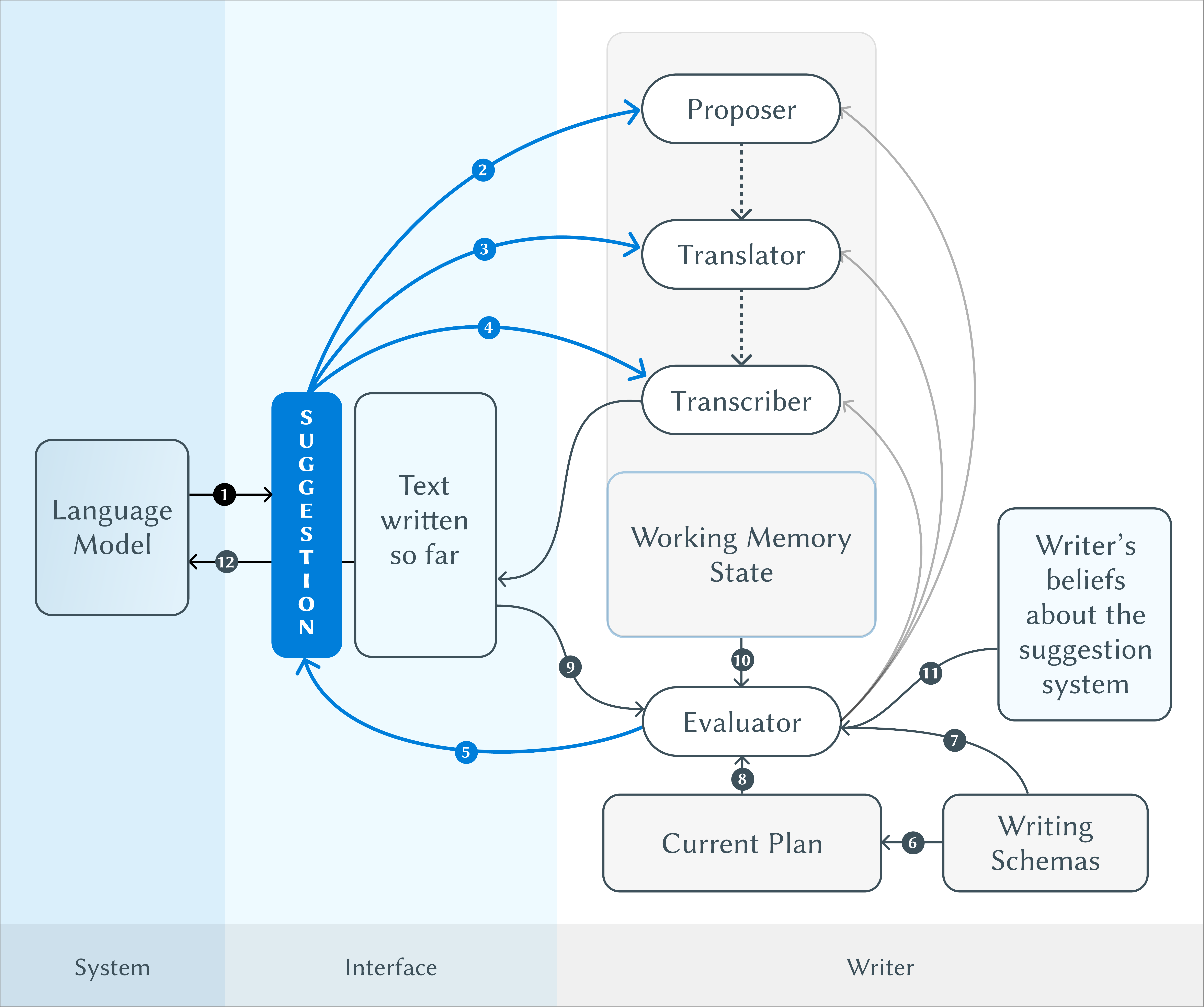}
  \caption{Our Writer-Suggestion Interaction model}
    \label{fig:our-model}
  \Description{A woman and a girl in white dresses sit in an open car.}
\end{figure*}
\\\
Past work on suggestion systems has conceptualised suggestions as 'transcription aids' whose goal is improving writing speed and preventing errors.\cite{kristensson2014inviscid} This 'transcription-centric' model of writer-suggestion interaction, as implied by past studies, assumes that writers have a concrete, translated proposal ready in mind when they encounter a suggestion, and the suggestion system merely helps them transcribe it. Our findings illustrate that is often not the case, as suggestions also affect the content of the writer's composition in various ways. This calls for a more holistic conceptualisation of writer-suggestion interaction.

The \citet{doi:10.1177/0741088312451260} CPMW provides a comprehensive model of the writing process through the cognitive processes of writing. The CPMW, through years of development, has also tried to accommodate and account for various factors that contribute to the writing process, such as collaborators, critics and transcription technologies. It, however, does not have a clear place for suggestion systems — which are neither passive transcription aids nor ‘collaborators or critics’. 
Our findings show that, at least in the context of our study, it is possible to look at writer-suggestion interaction as an interaction between the cognitive processes of writing and the suggestions generated by the language model. Based on our findings, we propose a model that builds upon the categories and concepts proposed by Hayes and articulates the findings of our study. The flow of information in the model starts when the writer encounters a suggestion (Fig \ref{fig:our-model}). The model consists of 3 parts: the writer, the interface and the language model.
\\[0.07in]
\emph{Language Model: }The language model (LM) sends suggestions to the interface (arrow 1) based on the text written so far by the writer in the writing interface. The LM, which in this case is GPT-2, is trained on a particular corpus (IMDb movie review) and may be aligned with the writer’s intent to a lesser or greater extent.
\\[0.07in]
\emph{Interface: }The interface is part of the task environment and consists of the text written so far and the suggestions presented through a specific interface design (refer to apparatus for the design of our writing interface).
\\[0.07in]
\emph{Writer: }As we described in our findings, a suggestion may contribute new ideas to the proposer (arrow 2), vocabulary and sentence constructs to the translator (arrow 3), and reduce typing and spelling effort, thereby contributing to the transcriber (arrow 4). We also observed that a single suggestion might contribute to a combination of these processes. This process of suggestions contributing to the processes of writing may happen subconsciously, where the writer (and the processes of writing) may get 'influenced' by the visible suggestion, or consciously, when a writer may aid their processes of writing with the help of a suggestion. We also define the writer's 'working memory state' as the state of their writing processes when they encounter and attend to a suggestion. This state may contain a partial or complete proposal, or a partial or complete translation.

Our model also borrows the 'evaluator' from the Hayes model, but evaluates the outcomes of the three writing processes along with the outcomes of the language model, i.e. the suggestions (arrow 5). Writers may evaluate the suggestions based on their writing schema and writing plans, as well as compare the suggestions with the text they have written so far (arrow 9). 

Writers also may compare their current 'working memory state' with the suggested text (arrow 10), check for consistency and whether the suggestion improves the partial or complete composition they have in mind. Writers may be less open to suggestions if they already have a concrete composition in mind.

Finally, the writers’ beliefs about how the suggestion system works may impact how open they are to the suggestions and how they evaluate them (arrow 11). For example, a writer who thinks the suggestion system reflects the views of other writers on the internet may be more willing to accept it.

The proposed model, to our knowledge, is the first attempt to articulate writer-suggestion interaction from a cognitive process lens. It is derived from qualitative analysis of a specific type of writing task - movie reviews, with a particular language model - GPT-2, fine-tuned on a specific data-set - IMDb movie reviews, with a one-dimensional bias (positive and negative sentiment) and using a particular suggestion interaction modality - single inline next-phrase suggestions. It can act as an analytical lens for interpreting qualitative and quantitative data, generating hypotheses, and providing vocabulary to articulate complex interactions observed between writers and suggestion systems (For example, interaction logs datasets from tools like CoAuthor~\cite{lee2022coauthor}). 

Further work on the model can include testing and validating the concepts and relationships proposed by the model, and theoretically expanding and generalising the model through empirical research. This may include conducting similar studies across different participant groups with different levels of language proficiency(L1 and L2 in English and regional languages), different writing tasks(Emails, Opinions, social media comments etc.), interaction modalities(multiple suggestions, smartphone keyboard etc.) and alignments(multi-dimensional opinions on a topic, misinformation etc.). Future work can also look at adapting the model in other domains, such as programming. In the past, verbal protocols have been used to study cognitive processes involved in programming~\cite{Brooks1999TowardsAT}. Similarly, there has been a recent interest~\cite{copilotstudy} in qualitatively studying how programmers interact with AI-assisted programming tools like Github Copilot~\cite{copi}. We believe future studies can use a similar methodology as ours to study user-suggestion interaction in domains such as programming from a cognitive process lens.

\subsection{Framing interaction with the writer as interaction with writing processes}
Our model conceptualises the written artefact composed with the aid of suggestions as the product of the interaction between the suggestions and the cognitive processes of writing. Hence, a suggestion system's effects on the cognitive processes of writing would manifest in the written product.

Prior work by Arnold et al. points out that the content of suggestions can affect the final written output. They showed that a positive bias in the suggestion system skews the sentiment of the final written product positively \cite{arnold2018sentiment}. With the help of our model, we can point out that it remains unclear whether this effect was at the language level or at the idea level, i.e. on the translator or the proposer. Future studies can benefit from this distinction to measure where the sentiment shift is taking place by choosing appropriate analysis strategies and different units of analysis. A comparison between topics suggested by the suggestions and the writer's topics could measure the influence on the proposer. In contrast, a language-level analysis of sentiment (\citet{arnold2018sentiment} ), vocabulary (\citet{arnold2020predictive}), or sentence structure etc., could be more suited for measuring the influence on the translator.

Making a distinction between the cognitive processes can also help in designing writing aids specifically targeted to particular writing processes. In a 2016 study, \citet{arnold2016suggesting} compared the impact of suggesting phrases vs words for writing restaurant reviews. They found that phrases aided writers more with ideas (what to say) and language (how to say it). Writers also reported that phrases provided them with more ideas than single-word suggestions. In a separate study, \citet{arnold2020predictive} also observed that suggesting single words led to writers using vocabulary presented by the suggestions instead of coming up with words themselves. Our findings echo this observation, with several participants attempting to align with genre-specific language and textual structure for the movie reviews. Studying how such design choices might aid or influence individual cognitive processes of writing can help develop design parameters that can be controlled to specifically aid the translator, proposer or transcriber. One such parameter could be the sampling strategy used for generating suggestions from the language model.

In our study, we observed that some writers found the suggestions generic; this echoes the findings of \citet{arnold2018sentiment}. This could be because the suggestions were generated using beam search, which optimises for the most likely next word to construct a phrase. Different sampling strategies like nucleus or temperature sampling could be used and manipulated to generate more divergent suggestions that may be more helpful in generating proposals, while strategies like beam search could be used for aiding translation and transcription.

The proposed model can be used to evaluate how well competing tools aid a specific writing process and without influencing the other processes. Future tools can allow writers to choose which cognitive process they want the suggestion system to support, and when. When the writers find themselves short on proposals, they could ask for innovative ideas that differ from what they have been writing; when they struggle to come up with suitable translations, writers could ask for suggestions more compliant with a particular style of writing (e.g. movie reviews) or simply alternative translations. Writers who have decided what they want to write could ask for fewer distractions and more transcription help. Providing writers with this choice and control can potentially make AI-enabled writing interfaces more transparent, explainable, and useful.

\subsection{The evaluator needs attention.}
\subsubsection{Evaluator fatigue}
In our model, we represent the evaluator as a cognitive process that, along with evaluating the outputs of the three processes of writing —  i.e. proposer, translator and transcriber — also follows the same function for evaluating suggestions. Unlike the other three processes of writing, the evaluator interrupts the cognitive processes to accept or reject their outputs. With suggestions in the picture, the evaluator has to do the same for the suggestions, along with comparing the outputs of the suggestion system and the cognitive processes to determine the best fit.

While suggestions aid the proposer, the translator and the transcriber, they do not aid the evaluator. Instead, they demand the writer to engage in more evaluation, which can lead to \emph{evaluator fatigue}. Arnold et al. \cite{arnold2018sentiment} suggest that writers often found next-word and next-phrase suggestions distracting, a phenomenon we also observed. A potential cause could be the increased load on the evaluator --- when writers already have a semi-formed composition in their working memory and are compelled to evaluate it against the encountered suggestion or evaluate the suggestion by itself, thus forgetting their semi-formed compositions. Such fatigue on the evaluator may lead to various effects such as writers leaving out certain critical criteria while evaluating suggestions, evaluating suggestions less critically over time or conforming with suggestions to save evaluation effort --- all of which can give way to unintended influence.
\subsubsection{Muddled Judgment}
We also found that writers did not have a clear understanding of the functioning of the suggestion system. They came up with theories, which led to unclear or false beliefs. In literature, such theories have been defined as algorithmic folk theories \cite{10.1145/3476046}. 
These beliefs directly affected how writers evaluated the suggestions and how open they were to changing their compositions and direction while writing. Some writers (Like U6) conformed to the suggestions despite the original misalignment because they thought the suggestions reflected the majority opinion on the internet about the specific movie being reviewed. They believed the AI had access to all the data on the internet and could query and source accurate and contextual data, like a search engine. This led to writers giving AI a sense of authority, perhaps because of their prior experience with and general perception of search engines, making them more open to suggestions. A similar attitude of giving AI authority based on analogising its underlying functionality with other technologies was noted by \citet{10.1145/3491102.3517533}. Our study confirms these findings in the context of next-phrase suggestion systems. False beliefs may persuade writers to be more open to and potentially conform to the suggestions.
\subsubsection{A Way Forward}
We believe future systems should find ways to aid the evaluator along with the other three processes: by decreasing the evaluation load and providing relevant and correct information so that writers' evaluation criteria are better informed.
Future suggestion systems could have a language model that is personalised to the writer's writing style based on their previous written content, and users could control how personalised they want their suggestions to be. A personalised model may need less evaluation from the writer’s end as most suggestions would already comply with their writing style — i.e. the model would less likely generate texts which the writer would reject. Further research could focus on whether personalisation decreases evaluator load and thus decreases evaluator fatigue.

There could be several ways of informing the writer’s evaluation criteria to help them make better decisions. The goal would be to give writers relevant and correct information so that they are better equipped to evaluate these suggestions. To begin with, providing writers with an explanation of how the suggestion system functions may help them build an accurate mental model, avoiding the ill effects of incorrect beliefs about the system. Writing interfaces could better reflect how suggestions are generated. Informing writers of the capabilities and limitations of the language models can help them evaluate the suggestions better. Solutions such as Model Cards proposed by \citet{mitchell2019model} can be helpful. These model cards could be simplified and made more user-facing. Attention visualisation techniques such as BertViz \cite{vig2019multiscale} may be implemented in text editors to highlight words written by the writer that contributed to the generation of the suggestion.

Helping the writers measure the impact a particular suggestion may have on their overall composition can also be useful. One way may be to give writers an overall sentiment score for their product. This score could be set to update as the text is written and when a suggestion is accepted, giving real-time feedback on how a suggestion impacts overall sentiment. A Grammarly-like approach to evaluating the tone of a piece of text may be put in place to evaluate the tone and effects of individual suggestions, along with an indication of which words/phrases contribute to the tone \cite{doi:10.1177/00336882211010506}.

\subsection{Limitations}

This study has some limitations. First, we generated suggestions after the writer paused for 300ms, against generating suggestions only when the confidence value is very high --- like in deployed systems. This decision helped us create more opportunities for observing writer-suggestion interaction. However, we acknowledge that showing suggestions more often would have affected how writers interacted with suggestions, and our findings would've been slightly different if the suggestions were not as frequent.

Second, we fine-tuned two GPT-2 language model instances on positive and negative movie reviews. To calculate a representative misalignment score, we assumed that the positive and negative corpora had a uniform distribution allowing us to approximate the Corpus Rating to be 2.5 and 8.5. The Corpus Rating values and the Degree of Misalignment scores should only be considered representative measures and not accurate ones. 

Third, due to limitations in time and budget, we could not consider an ‘unaligned’ condition, where writers would be given a suggestion system trained on the whole IMDb review corpus from ratings spanning from 1-10. The behaviour of such a system would've differed from the biased suggestion systems. In our pilot studies, we observed that such a system adapted to the writer's sentiment as the writer wrote the review and created higher chances of ‘aligned’ conditions. While we had several cases of aligned conditions, our design choice allowed us to also understand what happens when conditions are less aligned. Future work could study unaligned writer suggestion conditions and expand, validate or contest our models and findings.

Four, we selected movie review writing as our writing task. We have stated the reasons for doing so in the method section. However, we acknowledge that writing movie reviews in itself are an uncommon and relatively benign writing task with fewer implications for the writer. Other forms of writing in the real world can be more nuanced and sensitive, like writing personal emails or opinion pieces on political issues. This calls for further research using more multidimensional writer-suggestion alignments.

Five, our participants were L2 English speakers from India. Hence, the findings cannot be wholly generalised beyond this group (i.e. L2 English speakers in India). That being said, English remains an official language in India, and while our participants had a first language (or mother tongue) that was not English, English was a medium of instruction for their K-12 education. The results of this study should hence be interpreted, keeping that in mind. Future work can look at expanding and generalising these findings and the proposed model to populations with L1 and L2 speakers in other English-speaking and non-English-speaking countries, along with conducting this study in regional languages.

Finally, we would like to acknowledge that our findings and the proposed model are a result of a specific set of conditions. Care needs to be taken while generalising these findings across different writing tasks, language models, biases and interaction modalities. Like in all qualitative studies, the findings and the conclusions should be interpreted as provisional and contestable.
\section{Conclusion}
We conducted a qualitative study with 14 writers who wrote two movie reviews each, one with a system that offered next-phrase suggestions and one that did not. In the with-suggestions condition, authors were randomly given a system with a corpus rating of 2.5 or 8.5 out of 10. Writers had varying ratings before writing the review, thus resulting in a range of positive and negative alignments between the writer intent and the system. Analysis of the think-aloud and retrospective protocols collected from the users gave us insights into how the suggestions affected the different parts of the writing process. 

Our findings suggest that writers use suggestions in multiple, nuanced ways, even when they don’t directly accept the suggestions. Writers took aid from suggestions, directly or indirectly, on various levels, such as proposing (idea generation), translating (language), and transcribing (aid in typing) by abstracting and extracting parts of the suggestions. We also found various ways and criteria in which writers evaluated these suggestions and that writer-suggestion misalignment primarily impacted the evaluation process. We also describe how the presence and content of the suggestions had several effects on the writing process, such as a change in the writer’s overall plan, increased distraction along with the role of misalignment.

We propose a model of writer-suggestion interaction, ways in which the model can be used, and opportunities for future research and designs.

\begin{acks}
Thank you Lakshya, Saloni, Kartik and Amaan for all your help!
\end{acks}

\bibliographystyle{ACM-Reference-Format}
\bibliography{bibliography}

\appendix

\end{document}